\documentclass[aps,prl,onecolumn]{revtex4}
\usepackage{graphicx} 
\usepackage{dcolumn}  
\usepackage{bm}       

\def\Rb{{\bf R}}  \def\rb{{\bf r}}  \def\Eb{{\bf E}}  \def\Lb{{\bf L}}
      \def\Gb{{\bf G}}
  \def\ii{{\rm i}}  \def\Hb{{\bf H}}

\begin{document}

\title{Electromagnetic forces and torques in nanoparticles irradiated by a plane wave}
\author{F. J. {Garc\'{\i}a de Abajo}}
\affiliation{Centro Mixto CSIC-UPV/EHU and Donostia International
Physics Center\\ (DIPC), Apartado 1072, 20080 San Sebasti\'an,
Spain}

\begin{abstract}

Optical tweezers and optical lattices are making it possible to
control small particles by means of electromagnetic forces and
torques. In this context, a method is presented in this work to
calculate electromagnetic forces and torques for
arbitrarily-shaped objects in the presence of other objects
illuminated by a plane wave. The method is based upon an expansion
of the electromagnetic field in terms of multipoles around each
object, which are in turn used to derive forces and torques
analytically. The calculation of multipole coefficients are
obtained numerically by means of the boundary element method.
Results are presented for both spherical and non-spherical
objects.

\end{abstract}
\maketitle

\section{Introduction}

Electromagnetic forces produced by intense focused lasers acting
on small particles have recently found application in trapping and
manipulating small particles in optical tweezers
\cite{AD1987,C91_1,G03,NBX97,NRH01}. Also, optically-powered
rotors have been produced either by scattering of
elliptically-polarized light \cite{FNH98} or by using particles of
helical shape \cite{GO01}. These are the counterpart of similar
effects observed at the atomic and molecular levels, including
molecular quantum rotors and motors \cite{HHC02,AA01}.

In this work, we calculate electromagnetic torques acting on
elongated particles illuminated by a plane wave. We show that the
magnitude of these quantities, as well as the sign of the torque,
can be controlled by using the right polarization and wavelength
for the external light. We use a multipole formalism to calculate
both torques and forces, which can be applied to complex
geometries involving more than one particle, as illustrated below
for forces acting on metallic spheres in the presence of
neighboring particles of different shapes.

For practical applications in the context of optical tweezers
\cite{G03} and optical stretchers \cite{GAM00}, an extension of
the present study to include focused beams will be necessary,
similar to the one carried out in Ref. \cite{NRH01}, where
gradient forces are essential to achieve trapping. Even in this
case, one would expect that control over the orientation of small
particles is attainable by choosing the right combination of
polarization and wavelength. However, the present work can find
some relevance in different situations: (1) to control the
orientation of particles in free space by irradiating them with
successive plane-wave pulses of appropriate strength, duration,
wavelength, orientation, and polarization, specially if the
absolute spatial position is not so relevant; (2) to explore the
dynamics (both translational and rotational) of complex particles
in inter-stellar environments; or (3) to control the orientation
and position of particles trapped against a solid-fluid interface
(the analysis becomes straightforward if the dielectric constant
is approximately the same on either side of the interface),
although friction and other interfacial forces can play a
substantial role in this case.

The electromagnetic response of each of the particles considered
in this work has been expressed in terms of their corresponding
multipole $t$-matrix, which relates the coefficients of the
multipole expansion of the induced electromagnetic field to those
of the external field \cite{paper039042,paperhhh}. These
$t$-matrices are in turn obtained by solving Maxwell's equations
using the boundary element method \cite{paper027060,paperggg}.

\section{Torques on non-spherical particles}

We begin by expressing the electromagnetic field acting on a given
particle in terms of magnetic and electric multipoles in frequency
space $\omega$ as \cite{L97,paper039042,paperhhh}
   \begin{eqnarray}
      \Eb(\rb,\omega)=\sum_{lm} \{&&
          [\psi^{M,{\rm inc}}_{lm}\Lb
          -\psi^{E,{\rm inc}}_{lm}\frac{\ii}{k}\nabla\times\Lb] \ii^l j_l(k r)]
   \label{e1} \\
         +&&[\psi^{M,{\rm sc}}_{lm}\Lb
          -\psi^{E,{\rm sc}}_{lm}\frac{\ii}{k}\nabla\times\Lb] \ii^l h_l^{(+)}(k r)]\}
      Y_{lm}(\hat{\rb}),
   \nonumber
   \end{eqnarray}
where $\Lb=-\ii \rb\times\nabla$ is the orbital angular-momentum
operator, $k=\omega/c$ is the momentum of the light, and the field
has been separated in incident (inc) and scattered (sc)
components. Assuming linear response, the coefficients of
proportionality between them are given by the scattering matrix
$t$ according to
   \begin{eqnarray}
      \psi^{\nu,{\rm sc}}_{lm} =
      \sum_{l'm'} t^{\nu\nu'}_{lm,l'm'}\psi^{\nu',{\rm
      inc}}_{l'm'},
   \label{ggg}
   \end{eqnarray}
where $\nu$ runs over $M$ and $E$ components. The $t$ matrix is
analytical for spherical particles \cite{paper039042,paperhhh}
(Mie coefficients) and we have calculated it numerically using the
boundary element method \cite{paper027060,paperggg} for
arbitrarily-shaped objects (whiskers and torii in the examples
offered below).

\begin{figure*}
\centerline{\scalebox {0.5}{\includegraphics{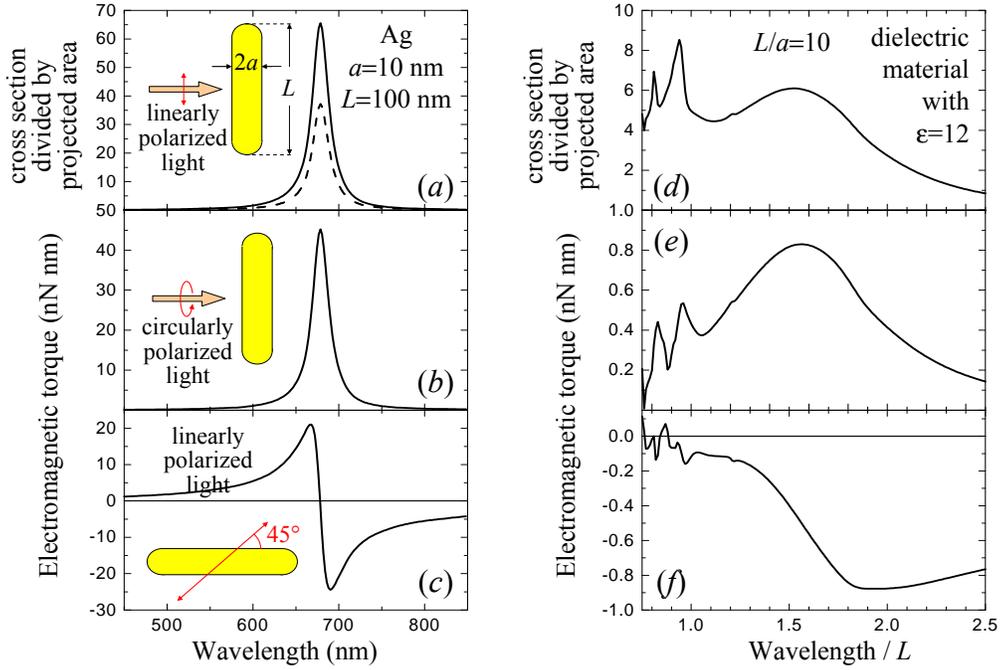}}}
\caption{{\textbf{(a)}} Scattering cross section for a silver
whisker as a function of wavelength. The shape and dimensions of the
axially-symmetric sample are shown in the inset. Both elastic
(dashed curve) and total (solid curve) cross sections are shown. The
light is linearly polarized along the axis of the whisker.
{\textbf{(b)}} Electromagnetic torque for the same whisker as in (a)
when it is illuminated by a circularly-polarized light plane wave of
intensity equal to 1 W/$\mu$m$^2$. {\textbf{(c)}} Electromagnetic
torque under linearly-polarized illumination with the polarization
vector forming an angle of 45 degrees with the whisker axis (the
light is coming along a direction perpendicular to the axis of the
whisker). {\textbf{(d)-(f)}} Same as (a)-(c) for a dielectric
whisker of dielectric function $\epsilon=12$.} \label{Fig1}
\end{figure*}

The torque acting on the particle in the presence of this field is
a quadratic, analytic function of these coefficients that can be
obtained from the integral of the Maxwell stress tensor over a
spherical surface $S$ of radius $R$ surrounding the object. The
time-averaged torque reads \cite{J1975}
\begin{eqnarray}
  \Gb= \frac{-R^3}{4\pi} \int_S d\hat{\Rb} [(\hat{\Rb}\cdot\Eb)
  (\Eb\times\hat{\Rb}) + (\hat{\Rb}\cdot\Hb)
  (\Hb\times\hat{\Rb})].
  \label{aaa}
\end{eqnarray}
Then, calculating the magnetic field $\Hb$ using Faraday's law and
inserting the resulting expression together with Eq.\ (\ref{e1})
into Eq.\ (\ref{aaa}), one finds
\begin{eqnarray}
  \Gb &=& \frac{1}{4\pi k^3} \sum_{lmm'} l (l+1) \;
  {\rm Re}\{[\frac{1}{2} \sqrt{(l+m+1)(l-m)} \delta_{m+1,m'} (\hat{\bf
x}-\ii\hat{\bf y}) \label{bbb} \\ &+& \;\;\;\;\; \frac{1}{2}
\sqrt{(l-m+1)(l+m)} \delta_{m-1,m'} (\hat{\bf x}+\ii\hat{\bf y}) + m
\delta_{mm'} \hat{\bf z}] \nonumber \\ &\times& [\psi^{E,{\rm
sc}}_{lm} (\psi^{E,{\rm sc}}_{lm'})^* + \psi^{M,{\rm sc}}_{lm}
(\psi^{M,{\rm sc}}_{lm'})^* + \ii \psi^{E,{\rm sc}}_{lm}
(\psi^{E,{\rm inc}}_{lm'})^* + \ii \psi^{M,{\rm sc}}_{lm}
(\psi^{M,{\rm inc}}_{lm'})^*]\}.
  \nonumber
\end{eqnarray}
Eq.\ (\ref{bbb}) has been used here to obtain the torque acting on
metallic and dielectric whiskers illuminated by a light plane
wave, as shown in Fig.\ \ref{Fig1}.

In the case of silver particles, the scattering cross section,
which is directly obtained from the scattering amplitude, shows a
pronounced resonance when the polarization of the external light
is directed along the whisker [Fig.\ \ref{Fig1}(a)]. The
dielectric function of silver has been taken from optical data
\cite{P1985}. Part of the scattered light is absorbed by the metal
(silver in this case), so that the total cross section (solid
curve, obtained by using the optical theorem \cite{L97}) is
actually larger than the elastic cross section (broken curve). The
cross section for polarization perpendicular to the particle is
negligible (by a factor of 40000 at the resonance) as compared
with the case considered in Fig.\ \ref{Fig1}(a).

The torque acting on this particle when it is illuminated by
circularly-polarized light follows the same profile as the
scattering cross section [Fig.\ \ref{Fig1}(b)]. This torque is
induced by the angular momentum carried by the external light,
part of which is transferred to the particle \cite{FNH98}.

A more interesting situation is presented when linearly-polarized
light is used [Fig.\ \ref{Fig1}(c)], in which case the particle
tends to align itself parallel (perpendicular) to the polarization
vector when light of wavelength below (above) the resonance is
employed. Here, the torque scales with the sine of the angle
between the polarization vector and the particle axis of symmetry.
It should be noted that the torque takes non-negligible values
well outside the absorption resonance, so that a sizable
orientational effect can still be obtained while minimizing heat
transfer to the particle (this arises from absorption of external
light). Moreover, control over the particle orientation is
possible by using the right combination of polarization and
wavelength of the external light.

It is interesting to point out that the wavelength of the
resonance depends on the length of the whisker, and this offers
the possibility of manipulating separately whiskers of different
lengths by tuning their respective resonances.

For dielectric particles of with same shape [Fig.\
\ref{Fig1}(d)-(f)] , the value of the torque is one order of
magnitude smaller and qualitatively very different as compared to
the metallic particles discussed above. For instance, dielectric
particles do not exhibit plasmon resonances, unlike metallic ones.
Moreover, their total and elastic cross sections are identical,
since a dielectric particle (real dielectric function) cannot
dissipate energy, so that heating of the particle is avoided.

\section{The effect of environment on electromagnetic forces}

For aggregates formed by several scattering objects, one can still
use the multipole expansion of Eq.\ (\ref{e1}) around each of the
objects. The scattered part of the self-consistent field around a
given object labeled $\alpha$ is the sum of contributions coming
from the other objects ($\beta\neq\alpha$) plus the scattering of
the incident field. Both scattering at each object and propagation
of the field between objects are linear operations (this would be
different in non-linear materials), so that the self-consistent
multipole coefficients satisfy the equation
  \begin{eqnarray}
      \psi^{\rm sc}_\alpha =
             t_\alpha (\psi_\alpha^{\rm inc} +
             \sum_{\beta\neq\alpha}
                          H_{\alpha\beta} \psi^{\rm sc}_\beta),
   \label{eq9}
   \end{eqnarray}
where matrix notation has been used, that is, $\psi$ is actually a
vector that contains all $M$, $E$, and $(l,m)$ components, and $t$
is the matrix of coefficients $t^{\nu\nu'}_{lm,l'm'}$, as defined
in Eq.\ (\ref{ggg}) . Here, the matrix $H_{\alpha\beta}$ describes
the propagation of the field from object $\beta$ to object
$\alpha$, and it can be derived analytically in terms of the
coordinates of the multipole origins for the different objects
\cite{paper039042,paperhhh}. Eq.\ (\ref{eq9}) separates the
geometrical configuration of the cluster, fully contained in
$H_{\alpha\beta}$, from the actual shape and composition of the
objects, which is entirely buried into $t_\alpha$. A similar
approach can be also followed to treat two-dimensional geometries
as well as photonic crystals consisting of periodic configurations
of the objects \cite{paper073074,paperjjj}.

\begin{figure*}
\centerline{\scalebox {0.5}{\includegraphics{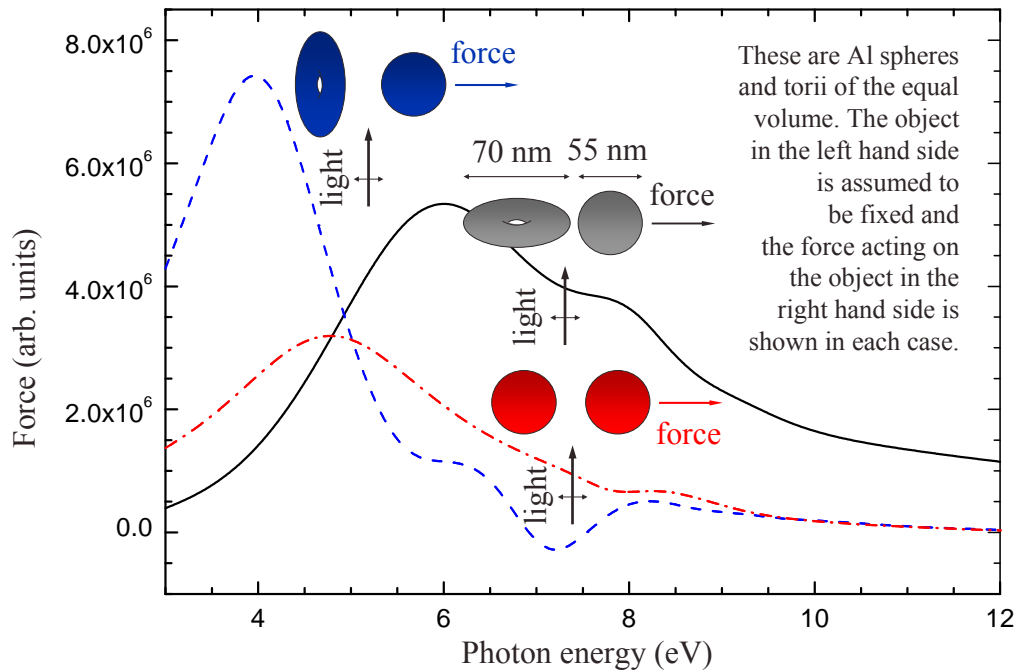}}}
\caption{Electromagnetic force acting on an aluminum sphere when it
is illuminated in the presence of a nearby object. The light is
linearly polarized, with the polarization vector contained in the
plane of the insets.} \label{Fig2}
\end{figure*}

This multiple scattering formalism has been used to obtain Fig.\
\ref{Fig2}, which shows the force acting on aluminum spheres of 55
nm in diameter when a nearby particle contributes as well to the
scattering of the external field. The electromagnetic force has
been calculated from the integral of Maxwell's stress tensor
\cite{J1975}, which results in an analytical but complicated
expression in terms of the multipole coefficients. A Drude
dielectric function has been used for aluminum, with a plasma
energy of 15 eV and a damping of 1.06 eV. The self-consistent
field has been obtained by using a method based upon multiple
scattering of multipoles \cite{paper039042,paperhhh}. A marked
influence of the neighboring particle is observed on the force
acting on the aluminum sphere, and the magnitude and even the sign
of this force changes dramatically over the photon energies under
consideration when choosing different particle shapes. This
suggests the possibility of using neighboring effects to control
the relative position of particles under the influence of external
light. The polarization of the latter has been chosen to maximize
these effects: a strong dipole-dipole interaction is triggered by
the component of the electric field directed along the line that
separates the centers of the objects, whereas the complementary
polarization results in minor neighboring effects that originate
in higher multiple contributions.

\section{Conclusions}

Torques and forces acting on small particles under external
illumination have been calculated in this work under illumination
by a single plane wave. Electromagnetic torques acting on
whiskers, both metallic and dielectric, have been shown to provide
a possible tool for nanoparticle alignment. The magnitude of the
torque for attainable light beam intensities is sufficiently large
as to overcome other forces such as gravity and Brownian motion.
Finally, the effect of neighboring particles on the
electromagnetic force acting on small aluminum spheres has proven
to be very large, suggesting a possible way to manipulate the
relative orientation of neighboring objects under external
illumination. The present study can be easily generalized to
account for illumination under focused beams, which will be needed
to discuss situations of practical interest in optical tweezers.

\section{Acknowledgments}

The author acknowledges help and support from the Basque
Departamento de Educaci\'{o}n, Universidades e Investigaci\'{o}n,
the University of the Basque Country UPV/EHU (contract No.
00206.215-13639/2001), and the Spanish Ministerio de Ciencia y
Tecnolog\'{\i}a (contract No. MAT2001-0946).

\end{document}